\documentclass[aps,prb,preprint]{revtex4-2}  
\usepackage[utf8]{inputenc}

\usepackage{amsmath,amsthm,amssymb,amsfonts,dsfont}
\usepackage{siunitx,microtype,hyperref}
\usepackage[swedish,english]{babel}

\usepackage{graphicx,booktabs,rotating,colortbl,xcolor}

\usepackage[caption=false]{subfig}
\usepackage{enumitem}

\newcommand{\Id}{\mathds{1}}
\newcommand{\R}{\mathbb{R}}
\newcommand{\E}{\mathcal{E}}

\newcommand{\Prob}{\mathbb{P}}

\newcommand{\x}{\boldsymbol{x}}
\newcommand{\y}{\boldsymbol{y}}

\newcommand{\ab}{\boldsymbol{a}}

\newcommand{\Ab}{\boldsymbol{A}}
\newcommand{\Mb}{\boldsymbol{M}}
\newcommand{\Ub}{\boldsymbol{U}}
\newcommand{\Sigb}{\boldsymbol{\Sigma}}
\newcommand{\Vb}{\boldsymbol{V}}
\newcommand{\Xb}{\boldsymbol{X}}
\newcommand{\Xbtilde}{\boldsymbol{\tilde X}}
\newcommand{\fb}{\boldsymbol{f}}
\newcommand{\fbtilde}{\boldsymbol{\tilde f}}
\newcommand{\fbhat}{\boldsymbol{\hat f}}

\newcommand{\wb}{\boldsymbol{w}}

\newcommand{\mub}{\boldsymbol{\mu}}
\newcommand{\muhb}{\boldsymbol{\hat\mu}}

\newcommand{\vb}{\boldsymbol{v}}
\newcommand{\s}{\boldsymbol{s}}
\newcommand{\tb}{\boldsymbol{t}}
\newcommand{\w}{\boldsymbol{w}}
\newcommand{\p}{\boldsymbol{p}}
\newcommand{\q}{\boldsymbol{q}}

\newcommand{\eps}{\boldsymbol{\varepsilon}}

\newcommand{\Abcal}{\boldsymbol{\mathcal{A}}}
\newcommand{\tildeAbcal}{\boldsymbol{\tilde{\mathcal{A}}}}

\DeclareMathOperator*{\argmin}{argmin}
\DeclareMathOperator*{\spn}{span}

\begin{document}

	\title{Dimensionality and Background Cancellation in Energy Selective X-Ray Imaging}

	\author{Fredrik Grönberg}
	\email{gronberg@mi.physics.kth.se}
	\affiliation{Physics of Medical Imaging, KTH Royal Institute of Technology, AlbaNova University Center, SE-106 91 Stockholm, Sweden}

	\author{Mats Persson}	
	\affiliation{\phantom{}Physics of Medical Imaging, KTH Royal Institute of Technology, AlbaNova University Center, SE-106 91 Stockholm, Sweden}

	\author{Hans Bornefalk}	
	\affiliation{\phantom{}\phantom{}Physics of Medical Imaging, KTH Royal Institute of Technology, AlbaNova University Center, SE-106 91 Stockholm, Sweden}

	\begin{abstract}
		\noindent
		\textbf{Purpose:} The set of linear attenuation coefficients that belong to materials in the human body is commonly assumed to be spanned by two basis functions in the range of clinical x-ray energies, even though there is evidence that the dimensionality of this set is greater than two. It has not yet been clear that the use of a third basis function could be beneficial in absence of contrast agents.		
		\\
		\\
		\textbf{Approach:} In this work, the choice of the number of basis functions used in the basis decomposition method is studied for the task of producing an image where a third material is separated from a background of two other materials, in a case where none of the materials have a K-edge in the range of considered x-ray energies (20-140 keV). The case of separating iron from mixtures of liver and adipose tissue is studied with a simulated phantom which incorporates random and realistic tissue variability. 
		\\
		\\
		\textbf{Results:} Inclusion of a third basis function improves the quantitative estimate of iron concentration by several orders of magnitude in terms of mean squared error in the resulting image.
		\\
		\\ 
		\textbf{Conclusions:} The inclusion of a third basis function in the basis decomposition is essential for the studied imaging task and could have potential application for quantitative estimation of iron concentration from material decomposed images.
	\end{abstract}

	\maketitle

\section{Introduction}\label{sec:intro}

	X-ray imaging techniques based on the basis decomposition method introduced by Alvarez and Macovski \cite{alvarez1976energy}, \cite{macovski1976energy} commonly assume that two basis functions span the set of linear attenuation coefficients (LACs) that belong to human tissue in the range of clinical x-ray energies (20-140 keV). This assumption is typically made with the argument of sufficient accuracy for the relevant application and its validity has been studied by several authors. In \cite{lehmann1986energy}, Alvarez presents the approximation error of body material LACs as a function of the number of basis functions used, and the accuracy of two-basis models is studied in detail by Gingold and Hasegawa in \cite{gingold1992systematic} and Willamson et al. in \cite{williamson2006two}. Gingold and Hasegawa point out in \cite{gingold1992systematic} that two basis functions are not enough to parametrize the set of LACs exactly and that this choice therefore leads to a systematic bias in quantitative measurements. In \cite{bornefalk2012xcom}, Bornefalk studies the intrinsic dimensionality of LAC data for low-Z elements ($Z = 1,\dots,20$) from the XCOM database \cite{berger1998xcom} and shows that the intrinsic dimensionality of the LAC data is equal to four with statistical significance. It is thus credible that additional information about an imaged LAC distribution becomes available by using more than two basis functions. Although the amount information is expected to be small, it may well be of use.

	On the other hand, Alvarez shows in \cite{alvarez2013dimensionality} that the Cram\`er-Rao lower bound of the variance of each estimated component of the decomposition increases with the number of basis functions used, if the basis functions are not orthogonal with respect to an inner product that is specific to each projection line. There is, in other words, under almost all circumstances a cost in terms of noise that comes with increasing the number of basis functions. The choice of the number of basis functions is thus associated with an information-noise trade-off that might or might not be beneficial depending on the application. The aim of this paper is to investigate the potential benefit of using a third basis function to try and separate a third material from a background of two other materials in the image processing stage, an operation referred to in this paper as \emph{background cancellation}. The rest of this introduction aims to give an overview of the paper and some of the choices made by the authors.
	
	Basis decomposition maps LACs to a space of the same dimension as the number of basis functions, referred to in this paper as the \emph{coefficient space}. The background cancellation operation can be described as mapping in coefficient space which maps a set of designated \emph{background material} coefficient vectors to zero. The question of dimensionality is of particular interest to the background cancellation task when bias due to error in the representation of the background materials is comparable to the signal sought in the produced image, because in a greater coefficient space one may potentially include some of this error in the kernel of the background cancellation mapping. This concept is studied in Section \ref{sec:separation} of this paper.

	To evaluate the performance of two- and three-dimensional background cancellation mappings and hence the benefit or disadvantage of using a three-basis model in this application, a simulated phantom was developed with two purposes in mind. First, to represent realistic biological variability of human tissue, and second, to provide an imaging task relevant to the background cancellation operation. The chosen case is the potential diagnosis of hemochromatosis, or liver iron-overload, with multi-bin spectral CT, with the imaging task of creating an image indicative of iron concentration by performing a background cancellation that removes liver and adipose tissue. This particular case was chosen for two reasons; compared to the constituent elements of liver and adipose tissue, iron has relatively high $Z$ and therefore has the potential not to lie in the range of liver and adipose tissue in a three-dimensional coefficient space; it was furthermore concluded in \cite{nielsen1992noninvasive} that dual energy CT showed poor sensitivity in determination of iron concentrations in the most diagnostically relevant range. It should be noted that the K-edge of iron at 7.11 keV lies below and outside the range of x-ray energies considered in this paper, the dimensionality of the problem is thus not affected by any discontinuity in the LACs. The construction of the phantom is described in Section \ref{sec:phantom}.

	Principal component analysis (PCA) \cite{jolliffe2005principal} is used to construct the basis functions used in the basis decomposition. This choice was made for several reasons. Unlike when using material basis functions, it is not possible (or at least very hard) to make a bad choice of basis functions with PCA, such as choosing two basis functions that are very similar. It was furthermore concluded in \cite{weaver1985attenuation} that principal component basis coefficients are more sensitive to differences in the elemental composition of body materials than the coefficients of photoelectric and Compton basis functions. Also, PCA is able of capturing effects that are not fully described by models or single material basis functions and therefore suitable to studies of dimensionality. \cite{lehmann1986energy}, \cite{bornefalk2012xcom}. The method of constructing basis functions with PCA is described in Section \ref{sec:construction}.

	Since the main focus of this paper is to study the effect of the choice of the number of basis functions rather than any specific reconstruction method, basis decomposition is performed without regularization and filtered back-projection (FBP) is used to produce images. The image reconstruction and the subsequent background cancellation are described in Section \ref{sec:reconstruction}. 

	A figure of merit used to evaluate the images is proposed in Section \ref{sec:merit}. The results are presented in Section \ref{sec:results} and Section \ref{sec:conclusion} concludes the paper.

	A preliminary study of the three-dimensional background cancellation task for a similar but less realistic phantom than the one developed in this paper was previously presented in \cite{gronberg2015third}.

\section{Materials}\label{sec:models}

	In this section we present the measurement model that is typically used to represent multi-bin spectral CT systems and the basis decomposition framework used in multi-bin spectral CT. The description is based on the formalism developed by Roessl and Herrmann in \cite{roessl2009cramer}. 
	
\subsection{Measurement Model}\label{sec:measurement}
	
	We assume that an object with a LAC distribution $\mu(\x,\E)$, where $\x$ denotes position and $\E$ denotes energy, is imaged using a multi-bin photon counting detector system with $N$ projection lines corresponding to a set of detector elements, each having $K$ energy bins. A general measurement model for such a system is that the number of counts in the $k$th energy bin of the $i$th projection line, denoted $\gamma_i$, is a Poisson random variable $Y_{ik}$ with expected value
	\begin{equation}\label{eq:model}		
		\lambda_{ik} = \int_\R w_{ik}(\E)\exp\left(-\int_{\gamma_i}\mu(\x,\E)\,ds\right)d\E + r_{ik},		
	\end{equation}
	where $w_{ik}(\E)$ describes the energy distribution and number of detected photons in the $k$th energy bin of the $i$th projection line in the case of an unattenuated incident beam. The additive term $r_{ik}$ describes counts that are due to electronic noise and photon scatter.

	The following model was assumed for $w_{ik}(\E)$ in \eqref{eq:model},
	\begin{equation}\label{eq:spectrum}
		w_{ik}(\E) = w_k(\E) = I_0\Phi(\E)D(\E)S_k(\E),
	\end{equation}
	where $I_0$ denotes the number of photons in the unattenuated beam, $\Phi(\E)$ the normalized x-ray spectrum, $D(\E)$ the detection efficiency of the detector and $S_k(\E)$ the energy response function of $k$th bin. If $R(\E,\E')$ denotes the detector energy response function, i.e. the probability of a detection with energy $\E'$ given an interaction with energy $\E$, and $T_0,\dots,T_K$ the bin edges, then
	\begin{equation}
		S_k(\E) = \int_{T_{k-1}}^{T_k}R(\E,\E')d\E'.
	\end{equation}

	For the simulation performed in this study, the incident photon number $I_0$ and x-ray spectrum $\Phi$ were generated using the x-ray tube model of Cranley et al. \cite{cranley1997catalogue}, assuming 120 kVp tungsten spectrum, $7^\circ$ anode angle and 6 mm aluminum filtration. The detection efficiency $D$ was assumed to be ideal and the detector response function $R(\E,\E') = \delta(\E - \E')$. The noise term $r_{ik}$ was assumed to be zero for all $i$ and $k$.

\subsection{Basis Decomposition}\label{sec:decomposition}

	Under the basis decomposition hypothesis, the LAC distribution $\mu(\boldsymbol{x},\E)$ may be decomposed into $L$ spatially dependent basis coefficient distributions $a_l(\x)$ and energy dependent basis functions $f_l(\E)$,
	\begin{equation}\label{eq:decomposition}
		\mu(\boldsymbol{x},\E) = \sum_{l=1}^L a_l(\x)f_l(\E).
	\end{equation}
	We will use the term \emph{basis coefficient} to refer to $a_l(\x)$ evaluated at any particular $\x$ and the term \emph{basis image} to refer to the spatially dependent function $a_l(\x)$ or its estimate. 

	For compactness of notation, we define
	\begin{equation}
	\begin{split}
		\ab(\x) &\triangleq (a_1(\x)),\dots,a_L(\x))^T, \\
		\fb(\E) &\triangleq (f_1(\E),\dots,f_L(\E))^T,
	\end{split}
	\end{equation}
	where the superscript $T$ denotes the vector transpose. We refer to $\ab(\x)$ evaluated at any particular $\x$ as a \emph{coefficient vector} and the spatially dependent function $\ab(\x)$ as \emph{basis images}. We also define the line integral of $a_l(\x)$ along the $i$th projection line as 
	\begin{equation}
		A_{il} \triangleq \int_{\gamma_i}a_l(\x)\,ds, 
	\end{equation}
	and the vector of such line integrals as
	\begin{equation}
		\Ab_i \triangleq (A_{i1},\dots,A_{iL})^T.
	\end{equation}
	
	Inserting \eqref{eq:decomposition} into \eqref{eq:model} yields the following parametrization of $\lambda_{ik}$,
	\begin{equation}\label{eq:param_model}		
		\lambda_{ik}(\Ab_i) = \int_\R w_{k}(\E)e^{-\Ab_i^T\fb(\E)}d\E + r_{ik}.
	\end{equation}
	Note that $\{A_{il}\}_{i=1}^N$ is the set of line integrals corresponding to a discrete Radon transform of the basis image $a_l(\x)$. 

	Let $\y_i = (y_{i1},\dots,y_{iK})$, where $y_{ik}$ is the observed counts in the $k$th energy bin of the $i$th projection line (i.e. an outcome of $(Y_{i1},\dots,Y_{iK})$). By assuming that $Y_{i1},\dots,Y_{iK}$ are \emph{independent} Poisson random variables, the likelihood of an observed set of counts in a single projection line is
	\begin{equation}
		\Prob(\y_i\,|\Ab_i) = \prod_{k=1}^K \frac{\lambda_{ik}(\Ab_i)^{y_{ik}}}{y_{ik}!}\exp\left(-\lambda_{ik}(\Ab_i)\right).
	\end{equation}
	Taking the negative logarithm and dropping terms that are constant with respect to $\Ab_i$ we form the following parametrized negative log-likelihood function for measurements in the $i$th projection line
	\begin{equation}
		L_i(\Ab_i) = \sum_{k=1}^K \lambda_{ik}(\Ab_i) - y_{ik}\log\lambda_{ik}(\Ab_i)
	\end{equation}
	and we define the maximum likelihood estimator of $\Ab_i$ as
	\begin{equation}\label{eq:ML}
		\Ab_i^* = \argmin_{\Ab_i}L_i(\Ab_i)
	\end{equation}
	Basis decomposition is performed by solving \eqref{eq:ML} for $i = 1,\dots,N$. This framework can be further extended to form a likelihood function for the basis images $a_l(\x)$ which may include regularization, as was done in e.g. \cite{long2012multi} and \cite{schirra2013statistical}.

\section{Methods}\label{sec:methods}	

\subsection{Background Cancellation}\label{sec:separation}

	Assume that the imaged LAC distribution $\mu(\x,\E)$ has the following representation as basis images in coefficient space
	\begin{equation}
		\ab(\x) = \alpha(\x)\ab_1(\x) + \beta(\x)\ab_2(\x) + \gamma(\x)\ab_3,
	\end{equation}
	where the basis images $\ab_1(\x)$ and $\ab_2(\x)$ correspond to two \emph{known} materials with a biological variability that is incorporated into their spatial dependence and the coefficient vector $\ab_3$ corresponds to a third material of some particular interest, whose distribution, $\gamma(\x)$, we are interested in estimating. The key point of modeling spatial variation of $\ab_1(\x)$ and $\ab_2(\x)$ is to capture the error made by modeling them with constant, standard representations, e.g. the ones found in ICRU-44 \cite{weaver1985attenuation}. The mixture coefficients satisfy
	\begin{equation}
		\alpha(\x) + \beta(\x) + \gamma(\x) = 1,
	\end{equation}
	for all $\x$ in the imaged object.

	We will consider the problem of estimating $\gamma(\x)$ in a two- or three-dimensional coefficient space using known quantities and an estimate of $\ab(\x)$ obtained by basis decomposition. We will now derive a form of $\ab(\x)$ from which it is simple to appreciate the estimation error in both cases.

	Let $\ab_1$ and $\ab_2$ denote standard representation coefficient vectors of the materials modeled with $\ab_1(\x)$ and $\ab_2(\x)$ and let 
	\begin{equation}
		\Delta\ab_1(\x) \triangleq \ab_1(\x) - \ab_1, \quad \Delta\ab_2(x) \triangleq \ab_2(\x) - \ab_2.
	\end{equation}
	Furthermore, let $\eps(\x)$ denote the estimation error in $\ab(\x)$ due to reconstruction, including the number of basis functions used in the basis decomposition, such that
	\begin{equation}
		\hat\ab(\x) \triangleq \ab(\x) + \eps(\x)
	\end{equation}
	is the estimate of $\ab(\x)$. We define the compound estimation error as
	\begin{equation}
		\Delta\ab(\x) = \alpha(\x)\Delta\ab_1(\x) + \beta(\x)\Delta\ab_2(\x) + \eps(\x)
	\end{equation}
	which includes both background material model error of and reconstruction error. It follows that
	\begin{equation}\label{eq:estimation}
		\hat\ab(\x) = \alpha(\x)\ab_1 + \beta(\x)\ab_2 + \gamma(\x)\ab_3 + \Delta\ab(\x).
	\end{equation}

	\subsubsection{Two-Dimensional Coefficient Space}

		To estimate $\gamma(\x)$ in a two-dimensional coefficient space, we will make the assumption that $\ab_3\notin\spn(\ab_1-\ab_2)$. There is then a vector $\p\in\ker([\ab_2-\ab_1]^T)$ such that $\p^T\ab_3 = 1$. Now, rewrite \eqref{eq:estimation} as
		\begin{equation}
			\begin{split}
				\hat\ab(\x) &= (\alpha(\x) + \beta(\x))\ab_1 + \beta(\x)(\ab_2 - \ab_1) \\
						&+ \gamma(\x)\ab_3 + \Delta\ab(\x) \\
						&= (1 - \gamma(\x))\ab_1 + \beta(\x)(\ab_2 - \ab_1) \\
						&+ \gamma(\x)\ab_3 + \Delta\ab(\x).
			\end{split}
		\end{equation}
		It follows that
		\begin{equation}
			\p^T\hat\ab(\x) = (1 - \gamma(\x))\p^T\ab_1 + \gamma(\x) + p^T\Delta\ab(\x)
		\end{equation}
		and thus that
		\begin{equation}
			\gamma(\x) = \frac{\p^T\left(\hat\ab(\x) - \ab_1 - \Delta\ab(\x)\right)}{1 - \p^T\ab_1}.
		\end{equation}
		An estimate of $\gamma(\x)$ can thus be constructed from $\hat\ab(\x)$, $\p$ and $\ab_1$ (or $\ab_2$, which would yield the same result),
		\begin{equation}\label{eq:gamma2d}
			\hat\gamma(\x) = \frac{\p^T(\hat\ab(\x) - \ab_1)}{1 - \p^T\ab_1}.
		\end{equation}
		The error of this estimator is
		\begin{equation}\label{eq:estimation_error_1}
			\hat\gamma(\x) - \gamma(\x) = \frac{\p^T\Delta\ab(\x)}{1 - \p^T\ab_1},
		\end{equation}
		which is equal to zero if the compound estimation error $\Delta\ab(\x)$ is equal to zero.

	\subsubsection{Three-Dimensional Coefficient Space}

		To estimate $\gamma(\x)$ in a three-dimensional coefficient space, we will make the stronger assumption that $\ab_3\notin\spn(\ab_1,\ab_2)$. It is necessary that the dimensionality of the set of LACs is greater than two for this assumption to hold for any $\ab_3$. There is then a vector $\q\in\ker([\ab_1\;\;\ab_2]^T)$ such that $\q^T\ab_3 = 1$. It follows from \eqref{eq:estimation} that
		\begin{equation}
			\q^T\hat\ab(\x) = \gamma(\x) + \q^T\Delta\ab(\x)
		\end{equation}
		and thus that
		\begin{equation}
			\gamma(\x) = \q^T(\hat\ab(\x)-\Delta\ab(\x)).
		\end{equation}
		An estimate of $\gamma(\x)$ can thus be constructed from the estimated $\hat\ab(\x)$ and $\q$,
		\begin{equation}\label{eq:gamma3d}
			\hat\gamma(\x) = \q^T\hat\ab(\x).
		\end{equation}
		The error of this estimator is		
		\begin{equation}\label{eq:estimation_error_2}
			\hat\gamma(\x) - \gamma(\x) = \q^T\Delta\ab(\x),
		\end{equation}
		which is equal to zero if the compound estimation error $\Delta\ab(\x)$ is equal to zero. 

	\subsubsection{Observations}
		What can we say about the estimation errors of $\gamma(\x)$ in both cases? In the three-dimensional case there is no magnification factor in the denominator, it might however be that the magnitude of $\q$ is greater than the one of $\p$. More importantly, in the two-dimensional case any part of $\Delta\ab(\x)$ that lies in $\spn(\ab_1-\ab_2)$ will not affect the estimation error of $\gamma(\x)$, whereas in the three-dimensional case this holds for any part of $\Delta\ab(\x)$ that lies in $\spn(\ab_1,\ab_2)$, which is a greater space. We also know that the part of $\Delta\ab(\x)$ that is due to reconstruction error will be greater in the three-dimensional case due to the increased number of basis functions, as was shown in \cite{alvarez2013dimensionality}.

\subsection{Phantom Construction}\label{sec:phantom}

	As mentioned in the introduction, the purpose of the simulated phantom is to test the performance of the background cancellation mappings \eqref{eq:gamma2d} and \eqref{eq:gamma3d} on an interesting task under realistic conditions of tissue variability. As stated in the ICRU-44: ``It is imperative that body-tissue compositions are not given the standing of physical constants and their reported variability is always taken into account.'' \cite{white1989tissue}

	\subsubsection{Tissue Data}
	With this purpose in mind, LAC data for liver and adipose tissue was created using their reported ranges of water, fat and protein content, found in the ICRU-44 \cite[Tab. 4.4]{white1989tissue} (rather than the constant, standard representation also found in the same reference). These ranges, along with the trace element content of each tissue, found in \cite[Tab. 4.6]{white1989tissue}, are presented in Tab. \ref{tab:content_ICRU}. The resulting variability is incorporated into the phantom by having each pixel map to a different tissue LAC realization. Since the reported ranges of tissue content correspond to multiple individuals, arguably, no single individual exhibits that much variability. As a whole, the phantom is therefore better thought of as a worst-case scenario of variability than a realistic representation of any single individual. It is however constructed in such a way that the local variation is small, and therefore more representative of single individuals in small regions of interest.
	\begin{table}[hptb]
		\centering
		\begin{tabular}{ l | c | c | c | p{6cm} }		
		 & Water [\%] & Fat[\%] & Protein [\%] & Trace elements [\%] \\
		\hline 
		Liver & 63.6 - 81.9 & 1.1 - 11.5 & 16-22 & 0.2 Na, 0.3 P, 0.3 S, 0.2 Cl, 0.3 K \\
		Adipose & 10.9 - 21.0 & 62 - 91 & 5 & 0.1 Na, 0.1 S, 0.1 Cl \\		
		\end{tabular}
		\caption{Ranges of water, fat, protein and trace element content in weight fraction (\%) of liver and adipose tissue. \cite{white1989tissue}}\label{tab:content_ICRU}
	\end{table}
	In order to translate water, fat and protein content to elemental content, the elemental compositions of water, fat and protein found in the ICRP Report 23 \cite{richmond1985icrp} were used. These are presented in Tab. \ref{tab:content_ICRP}. Water was assumed to have a density of 1.00 $\text{g}/\text{cm}^3$, protein to have a density of 1.35 $\text{g}/\text{cm}^3$ \cite{fischer2004average} and fat to have a density of 0.87 $\text{g}/\text{cm}^3$ (a value of 0.9 $\text{g}/\text{cm}^3$ is found in \cite{durnin1974body}). 

	\begin{table}[hptb]
		\centering
		\begin{tabular}{ l | c | c | c | c }		
		 & H [\%] & C [\%] & N [\%] & O [\%]\\
		\hline 
		Water & 11 & 0 & 0 & 89 \\
		Fat & 12 & 77 & 0 & 11 \\
		Protein & 7 & 52 & 16 & 23
		\end{tabular}
		\caption{Elemental composition in weight fraction (\%) of water, fat and protein. \cite{richmond1985icrp}}\label{tab:content_ICRP}
	\end{table}

	\subsubsection{Tissue Model}

\begin{figure}[htbp]
		\centering
		\begin{minipage}[b]{\linewidth}
			\centering
			\subfloat[Liver content.]{\includegraphics[width=0.45\linewidth]{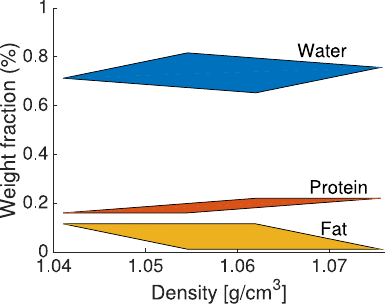}}				
			\hfil
			\subfloat[Adipose content.]{\includegraphics[width=0.45\linewidth]{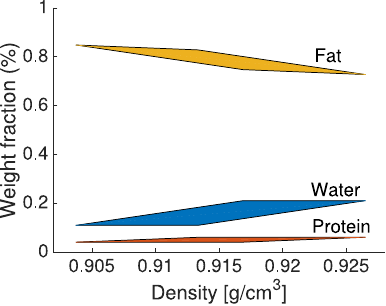}}			
		\end{minipage}
		\caption{Sets of valid ranges of water, fat and protein content in liver and adipose tissue.}\label{fig:tissue_sets}
	\end{figure}

	A tissue model based on ranges of water, fat and protein content can be formulated using the assumption that they form an incompressible mixture. Let $x$, $y$, $z$ denote weight fractions of three materials with corresponding densities $\rho_x < \rho_y < \rho_z$, and let $\xi$ denote the weight fraction of a material with unknown density $\rho_\xi$. Furthermore, let $l_x$ and $u_x$ denote the lower and upper bound of the range of $x$, and similarly for $y$ and $z$. In our case it will be that $x$ corresponds to fat, $y$ to water, $z$ to protein and $\xi$ to trace elements.	Since $\xi$ is small for both tissues we are simulating, we will assume that $\rho_\xi$ is constant and equal to one for both tissues, producing a small and presumably negligible error. A valid mixture satisfies the following set of constraints

	\begin{equation}\label{eq:tissue}
		\begin{split}			
			x + y + z &= 1 - \xi, \\
			l_x \leq x &\leq u_x, \\
			l_y \leq y &\leq u_y, \\
			l_z \leq z &\leq u_z,
		\end{split}
	\end{equation}
	which describes a convex, two-dimensional subset of $\R^3$. We wish to parametrize this set with two parameters $s$ and $t$. Our approach will be to parametrize the density with one parameter and parametrize all valid mixtures with a particular density with the other parameter. The density of a mixture is given by
	\begin{equation}
		\rho = x\rho_x + y\rho_y + z\rho_z + \xi.
	\end{equation}
	We begin by finding the fractions that correspond to a minimal and maximal density. Let $x_*$, $y_*$ and $z_*$ satisfy \eqref{eq:tissue} such that $\rho = \rho_{\min}$ is minimal and let $x^*$, $y^*$ and $z^*$ satisfy \eqref{eq:tissue} such that $\rho = \rho_{\max}$ is maximal. It can be found that
	\begin{equation}
		\begin{split}
			x_* &= \min(u_x,1 - \xi - l_y - l_z), \\
			y_* &= \min(u_y,1 - \xi - l_z - x_*), \\
			z_* &= 1 - \xi - x_* - y_*,
		\end{split}
	\end{equation}
	and
	\begin{equation}
		\begin{split}
			z^* &= \min(u_z,1 - \xi - l_x - l_y), \\
			y^* &= \min(u_y,1 - \xi - l_x - z^*), \\
			x^* &= 1 - \xi - y^* - z^*.
		\end{split}
	\end{equation}
	Let $\x_* = (x_*,y_*,z_*)^T$ and $\x^* = (x^*,y^*,z^*)^T$. It follows that
	\begin{equation}\label{eq:solution}
		\x_* + s(\x^* - \x_*),\quad s \in [0,1] 
	\end{equation}
	are solutions of \eqref{eq:tissue} with corresponding densities
	\begin{equation}
		\rho_{\min} + s(\rho_{\max} - \rho_{\min}).
	\end{equation}
	A valid mixture with density $\rho$ will satisfy the equality constraints
	\begin{equation}\label{eq:equality}
		\begin{split}
			x\rho_x + y\rho_y + z\rho_z &= \rho - \xi, \\
			x + y + z &= 1 - \xi,
		\end{split}
	\end{equation}	
	and all such valid mixtures can be found by adding an element
	\begin{equation}
		\vb\in\ker\left(\begin{bmatrix} \rho_x & \rho_y & \rho_z \\ 1 & 1 & 1 \end{bmatrix}\right), \quad \text{e.g.} \quad
		\vb = 
		\begin{pmatrix} 
			\rho_z - \rho_y \\
			\rho_x - \rho_z \\
			\rho_y - \rho_x
		\end{pmatrix},
	\end{equation}	
	to such a solution, for instance of the type \eqref{eq:solution}. Let $v_*(s)$ and $v^*(s)$ be the smallest, respectively greatest, multiple of $\vb$ that can be added to a solution of type \eqref{eq:solution}, without breaking any inequality constraints of \eqref{eq:tissue}. It then holds that 
	\begin{equation}\label{eq:parametrized}
		\begin{split}
			&\x(s,t) = \\
			&\quad\x_* + s(\x^* - \x_*) + \vb(v_*(s) + t(v^*(s) - v_*(s))), \\
			&\qquad s \in [0,1], \quad t \in [0,1]
		\end{split}
	\end{equation}
	is a parametrization of all solutions to \eqref{eq:tissue}. All that remains is to determine $v_*(s)$ and $v^*(s)$, which we will do by considering the extreme values of $t$. The signs of the components of $\vb$ tell us that adding a negative multiple of $\vb$ corresponds to addition of material $y$ and subtraction of materials $x$ and $z$. The bounds that have to be satisfied for $t = 0$ are thus $l_x$, $u_y$ and $l_z$. For $t = 0$, we thus have that
	\begin{equation}
	\begin{split}
		x_* + s(x^* - x_*) &+ (\rho_z - \rho_y)v_*(s) \geq l_x, \\
		y_* + s(y^* - y_*) &+ (\rho_x - \rho_z)v_*(s) \leq u_y, \\
		z_* + s(z^* - z_*) &+ (\rho_y - \rho_x)v_*(s) \geq l_z.
	\end{split}
	\end{equation}
	The smallest possible $v_*(s)$ that satisfies these inequalities is given by
	\begin{equation}
		v_*(s) = \max\left\{
		\begin{aligned}
		&\frac{l_x - x_* - s(x^* - x_*)}{\rho_z - \rho_y}, \\
		&\frac{u_y - y_* - s(y^* - y_*)}{\rho_x - \rho_z}, \\ 
		&\frac{l_z - z_* - s(z^* - z_*)}{\rho_y - \rho_x}.
		\end{aligned}
		\right.
	\end{equation}
	Vice versa, addition of a positive multiple of $\vb$ corresponds to addition of materials $x$ and $z$ and subtraction of material $y$. The bounds that have to be satisfied for $t = 1$ are thus $u_x$, $l_y$ and $u_z$. For $t = 1$, we thus have that
	\begin{equation}
	\begin{split}
		x_* + s(x^* - x_*) &+ (\rho_z - \rho_y)v^*(s) \leq u_x, \\
		y_* + s(y^* - y_*) &+ (\rho_x - \rho_z)v^*(s) \geq l_y, \\
		z_* + s(z^* - z_*) &+ (\rho_y - \rho_x)v^*(s) \leq u_z.
	\end{split}
	\end{equation}
	The greatest possible $v^*(s)$ that satisfies these inequalities is given by
	\begin{equation}
		v^*(s) = \min\left\{
		\begin{aligned}
		&\frac{u_x - x_* - s(x^* - x_*)}{\rho_z - \rho_y}, \\
		&\frac{l_y - y_* - s(y^* - y_*)}{\rho_x - \rho_z}, \\ 
		&\frac{u_z - z_* - s(z^* - z_*)}{\rho_y - \rho_x}.
		\end{aligned}
		\right.
	\end{equation}
	The ranges of water, fat and protein content used to simulate liver and adipose tissue are as presented in Tab. \ref{tab:content_ICRU}, except for the protein content of adipose tissue, which was allowed to be in the range of 4-6\%. The sets of valid content ranges are shown for both tissues in Fig. \ref{fig:tissue_sets}. The resulting density ranges for both tissues are similar to the ones found in \cite{white1989tissue}.

	\subsubsection{LAC Model}

	Let 
	\begin{equation}
		\mub_{\text{liver}}(s,t) \quad  \text{and} \quad \mub_{\text{adipose}}(s,t), \quad s\in[0,1],\quad t\in[0,1],
	\end{equation}
	denote the LAC data obtained by mixing elemental LAC data from the XCOM database \cite{berger1998xcom} according to the composition and density obtained from the parametrization \eqref{eq:parametrized} for each tissue. Water, fat and protein content are translated into elemental compositions using the values presented in Tab. \ref{tab:content_ICRP} and the trace element content presented in Tab. \ref{tab:content_ICRU} is included for each tissue. Let $\mub_{\text{iron}}$ denote the LAC of iron, also obtained from \cite{berger1998xcom}.

	\subsubsection{Phantom}

\begin{figure}[htbp]
		\centering
		\begin{minipage}[b]{\linewidth}
			\centering
			\subfloat[Example of the mixing texture $\w$. \label{fig:texture}]{\includegraphics[width=0.45\linewidth]{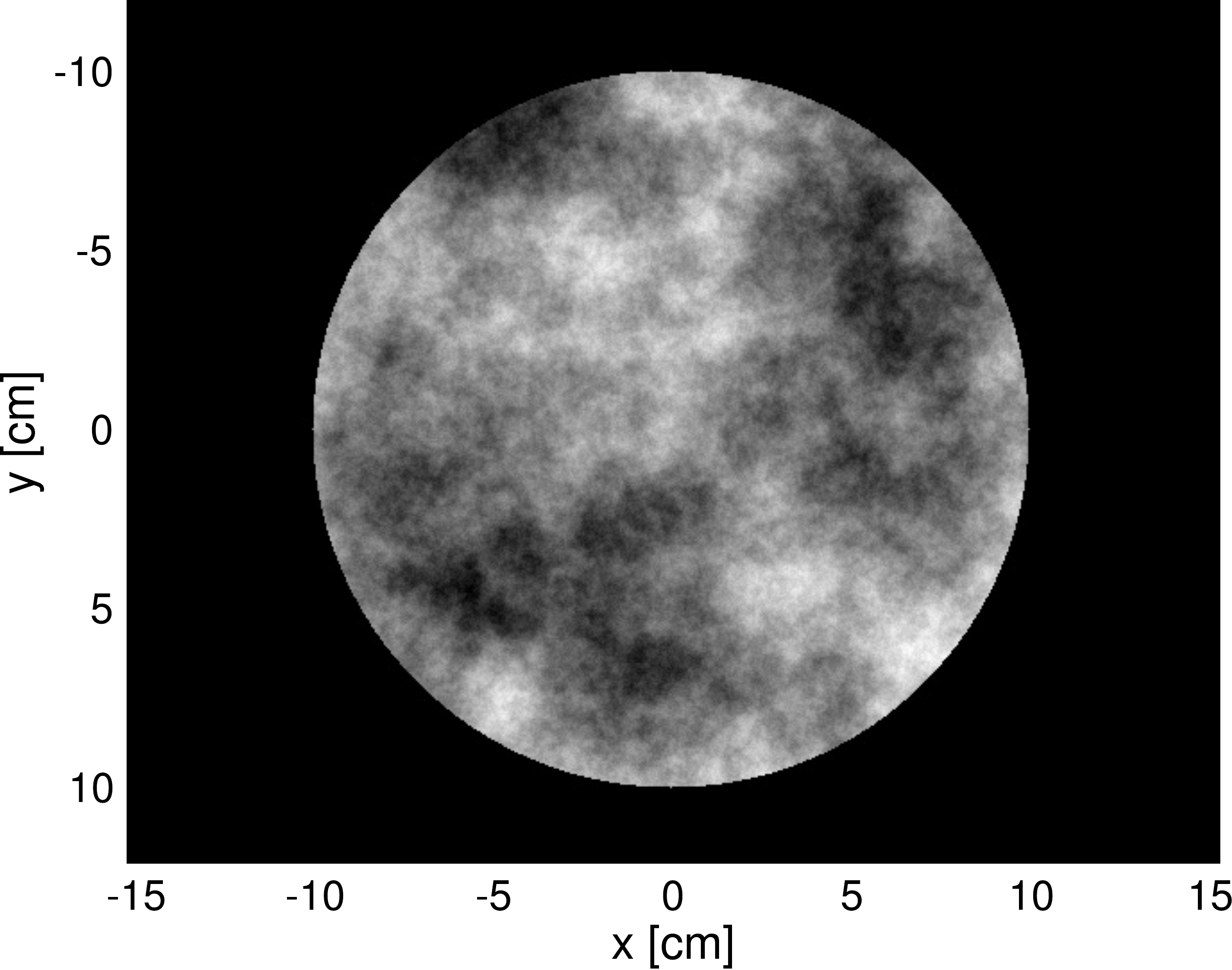}}				
			\hfil
			\subfloat[Central slice of the 2D NPS of the textures. \label{fig:nps}]{\includegraphics[width=0.45\linewidth]{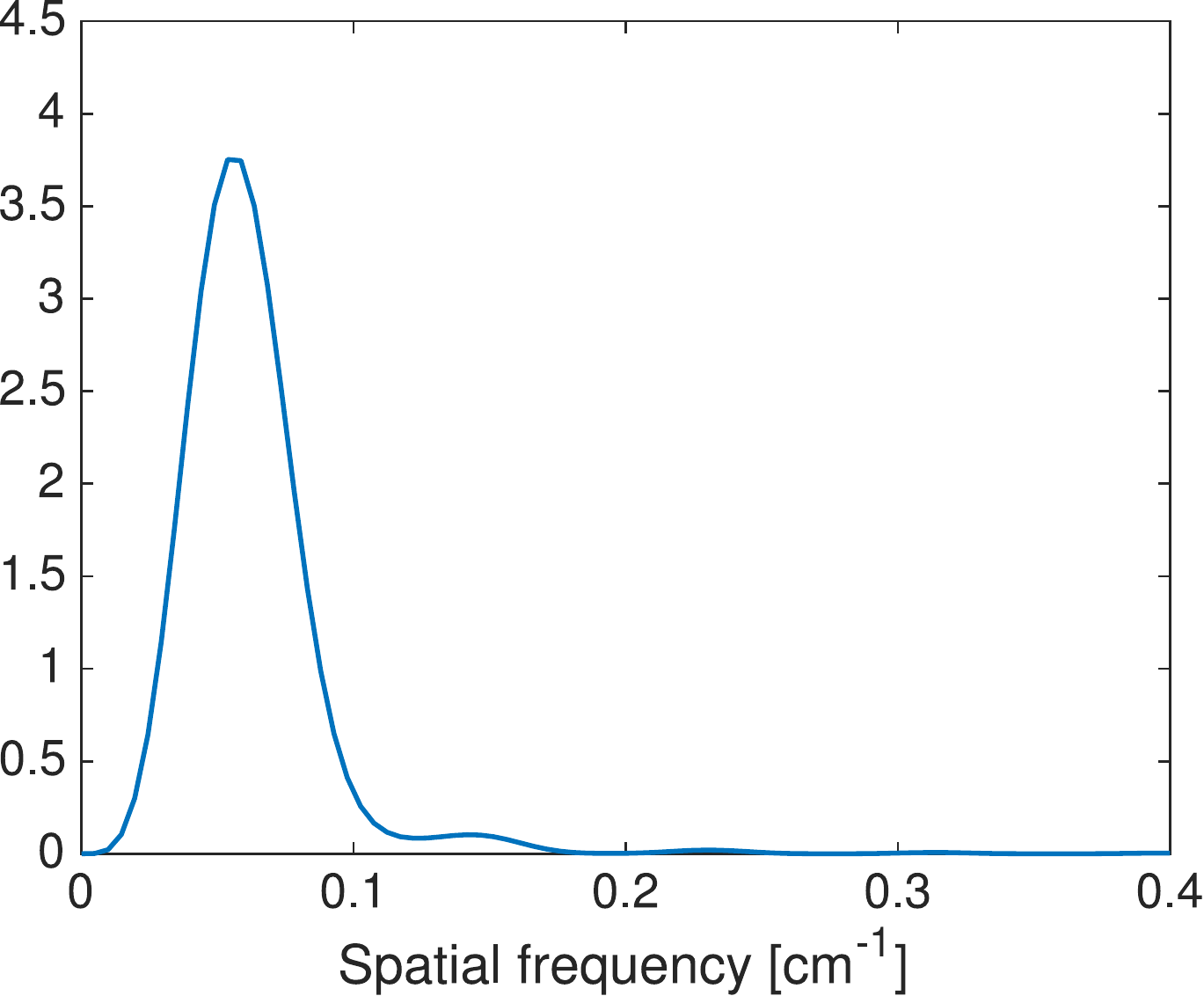}}
			\hfil		
			\subfloat[Iron weight map in logarithmic scale. \label{fig:iron_map}]{\includegraphics[width=0.45\linewidth]{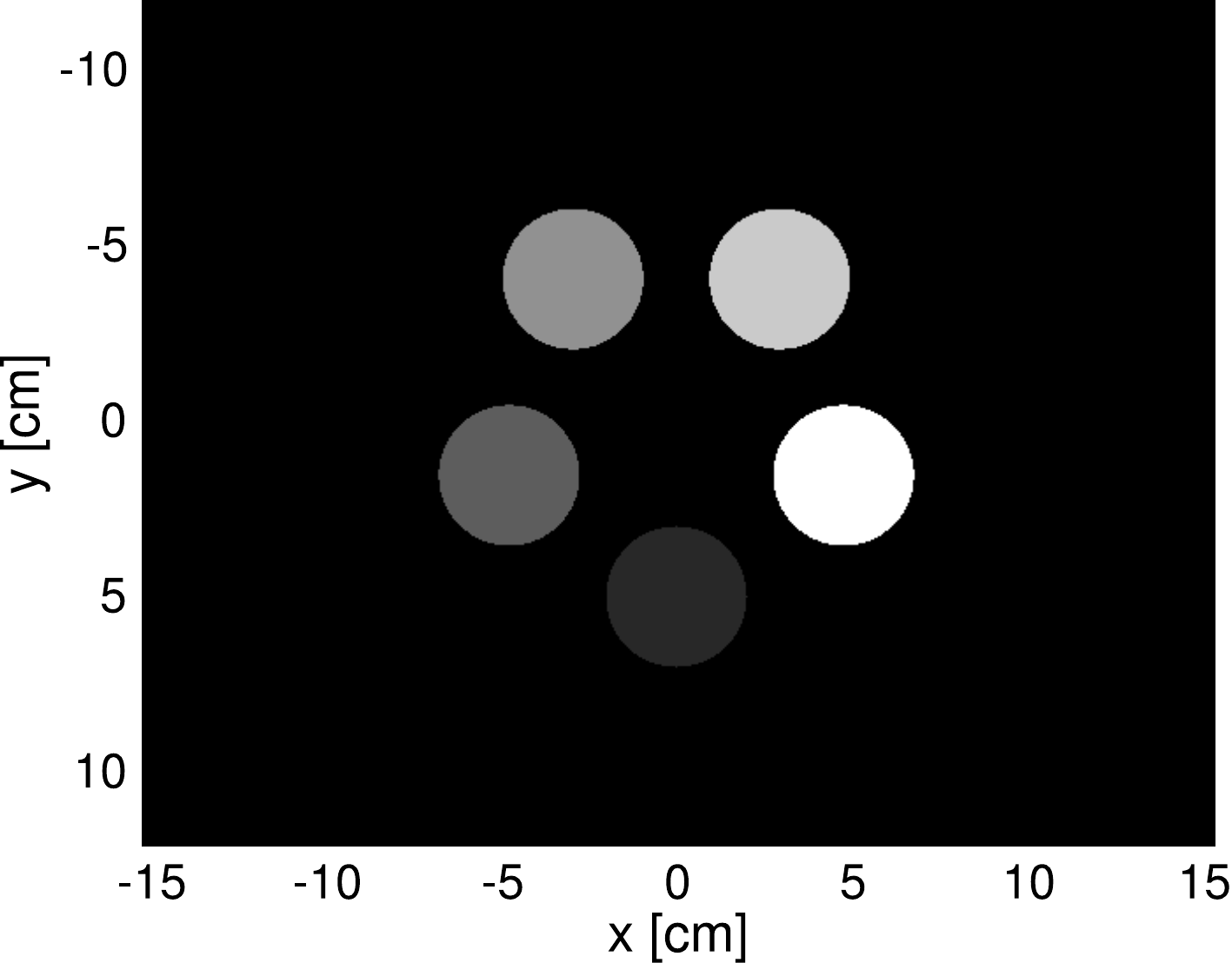}}
			\hfil
			\subfloat[Example of the phantom evaluated at $\E = 70$ keV, two of the inserts are to weak to distinguish from the background. \label{fig:phantom}]{\includegraphics[width=0.45\linewidth]{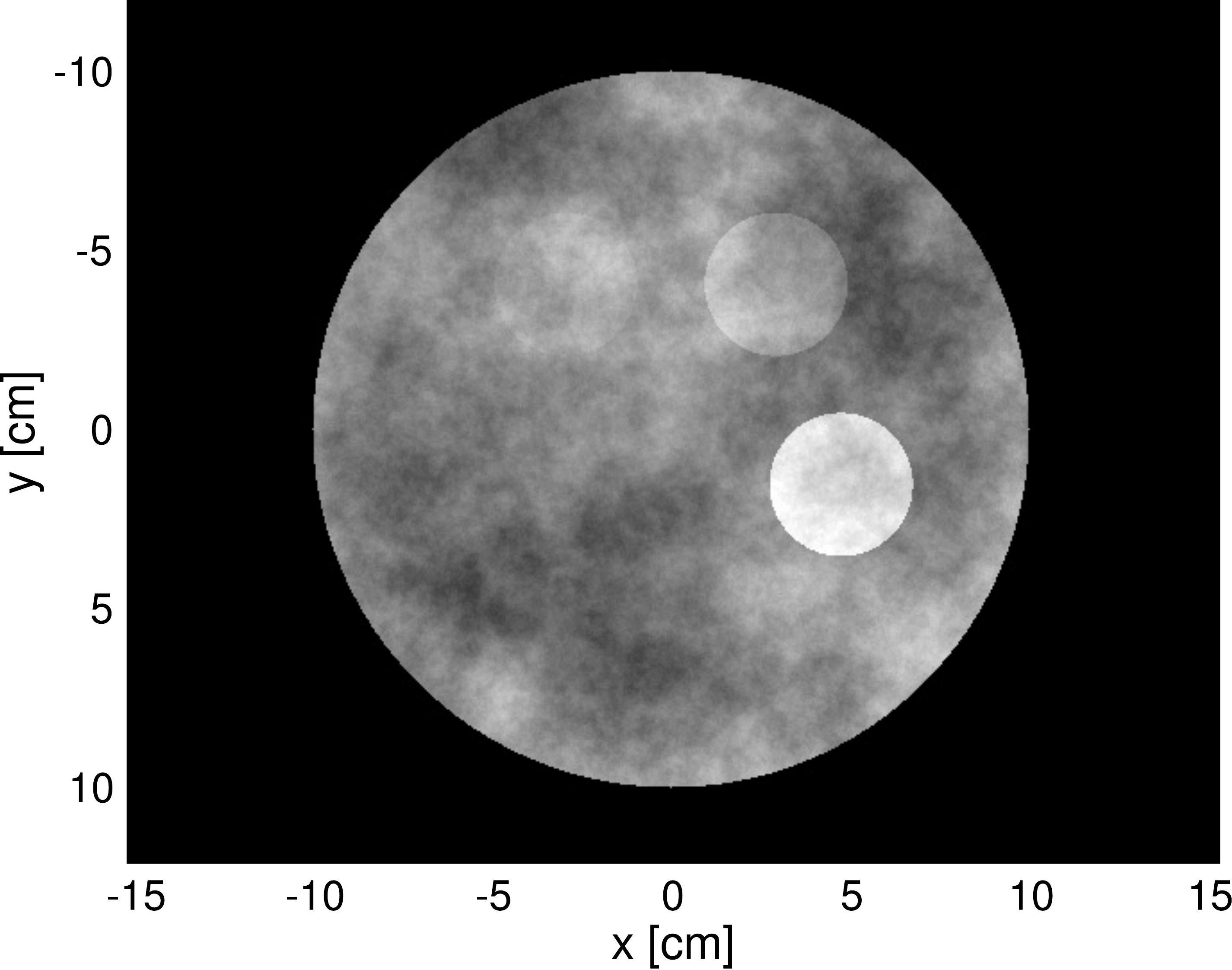}}
		\end{minipage}
		\caption{Phantom figures.}\label{fig:phantom_texture}
	\end{figure}

	To create the phantom, five two-dimensional random textures are generated using the diamond-square algorithm of Miller \cite{miller1986definition}: a mixing texture $\wb$, and material parameter textures $\s^{(k)}$ and $\tb^{(k)}$ for both tissues ($k=1,2$). The textures are offset and scaled such that $w_{ij}$, $s^{(k)}_{ij}$, $t^{(k)}_{ij} \in [0,1]$ for all $i$, $j$ and $k$. An example texture is shown in Fig. \ref{fig:texture}. The central slice of the textures’ noise power spectrum, which describes the frequency content of the phantom, is shown in Fig. \ref{fig:nps}.

	A two-dimensional phantom -- a 20 cm diameter cylinder with a pixel size of $0.5 \times 0.5$ $\text{mm}^2$, containing a random mixture of liver and adipose tissue along with five cylindrical inserts with iron concentrations of $(1/3, 1, 3, 9, 27)$ $\text{mg}/\text{cm}^3$, ranging from levels of normal iron stores to severe overload \cite{nielsen1992noninvasive} -- is created as follows
	\begin{equation}\label{eq:phantom}
	\begin{split}
		\mub_{ij} &= (w_{ij} - \gamma_{ij}/2)\mub_{\text{liver}}(s_{ij}^{(1)},t_{ij}^{(1)}) \\
			  &+ (1 - w_{ij} - \gamma_{ij}/2)\mub_{\text{adipose}}(s_{ij}^{(2)},t_{ij}^{(2)})\\
			  &+ \gamma_{ij}\mub_{\text{iron}},
	\end{split}
  	\end{equation}
  	where $\gamma_{ij}$ is the normalized iron concentration in pixel $(i,j)$ (w.r.t the concentration of 7.874 $\text{g}/\text{cm}^3$ used to compute $\mub_{\text{iron}}$). The iron weight map is shown, in logarithmic scale, in Fig. \ref{fig:iron_map} and an image of the constructed phantom, evaluated at E = 70 keV, is shown in Fig. \ref{fig:phantom}.

  	Poisson distributed projection data is created from this phantom using discretizations of the forward model \eqref{eq:model} and the spectrum model \eqref{eq:spectrum}, in integer keV steps. The detector elements are assumed to  be of size $1\times 1$ $\text{mm}^2$, and the system to have a source-to-detector distance of 100 cm and source-to-isocenter distance of 50 cm, corresponding to a spatial resolution of $0.5 \times 0.5$ $\text{mm}^2$ in isocenter. Assuming a current-time product of 300 mAs then yields $I_0 = 2.15 \cdot 10^5$ photons per projection line. Five energy bins are used and the bin edges are set to produce an approximately equal number of counts in each bin. Data is generated for 20 detector slices and added together, corresponding to a 10 mm slice in isocenter. 

\subsection{Basis Construction using PCA}\label{sec:construction}

	\subsubsection{Observations on Mean Centering}
	A common pre-computation step of PCA is mean centering and scale normalization of the features in the data set it is applied to. Mean centering is necessary to obtain an optimal approximation of the data set \cite{miranda2008new} and the purpose of rescaling is to not make any one feature dominate the resulting principal components. 

	In the context of constructing basis functions for the LAC, the mean component of the LAC is highly relevant to the forward model \eqref{eq:model}.	We therefore want to add a mean component back to each basis function after the application of PCA and dimensionality reduction. We first show that this is consistent with the basis decomposition framework and then how it is done in practice. From this section and forward, when referring to a \emph{basis vector}, what is implied is a discretized basis function. 

	Let $\langle\cdot\rangle$ denote the mean operator in the energy variable and let
	\begin{equation}
		\hat\mu \triangleq \langle\mu(\E)\rangle, \quad \hat f_l \triangleq \langle f_l(\E) \rangle,
	\end{equation}
	denote the energy means of a LAC $\mu(\E)$ and the $l$th basis function $f_l(\E)$ respectively. Let also
	\begin{equation}
		\tilde f_l(\E) \triangleq f_l(\E) - \hat f_l
	\end{equation} 
	denote the centered $l$th basis function. By \eqref{eq:decomposition} it follows that
	\begin{equation}\label{eq:mean}
		\hat\mu = \sum_{l=1}^L a_l\langle f_l(\E) \rangle = \sum_{l=1}^L a_l\hat f_l,
	\end{equation}
	Subtraction of \eqref{eq:mean} from \eqref{eq:decomposition} then yields that
	\begin{equation}\label{eq:centered}
		\mu(\E) - \hat\mu = \sum_{l=1}^L a_l(f_l(\E)-\hat f_l) = \sum_{l=1}^L a_l\tilde f_l(\E).
	\end{equation}
	Since $f_l(\E) = \hat f_l + \tilde f_l(\E)$ the decomposition of $\mu(\E)$ may also be expressed as
	\begin{equation}\label{eq:orig}
		\mu(E) = \sum_{l=1}^L a_k(\hat f_l + \tilde f_l(\E)).
	\end{equation}
	The argument is now as follows: By \eqref{eq:orig}, the non-centered $\mu(\E)$ may be decomposed with basis coefficients $a_l$ and centered basis functions $\tilde f_l(\E)$ found by performing e.g. PCA on a data set of centered LACs of the form \eqref{eq:centered}; and basis means $\hat f_l$ found by solving the corresponding system of basis coefficients $a_l$ and the set of LACs means of the form \eqref{eq:mean}. 

	\subsubsection{Basis Construction from LAC Data}
	Let $\mub_1,\dots,\mub_n\in\R^n$ be the set of LACs of \emph{air, water, iron, adipose tissue, whole blood, cortical bone, brain, breast tissue, eye lens, liver, lung tissue, skeletal muscle, testis and soft tissue}, sampled in the energy range $20-140$ keV in $n$ steps. Air, water and iron LACs were obtained from the XCOM database \cite{berger1998xcom} and body material LACs were computed by mixing LAC data of the elements $Z = 1,\dots,20,26$ from the XCOM database \cite{berger1998xcom} according to the compositions and densities found in ICRU-44 \cite{white1989tissue}. 

	We collect this data in a matrix $\Xb = [\mub_1,\dots,\mub_p] \in \R^{n\times p}$. The centered and scaled data matrix is defined as $\Xbtilde = [\frac{\mub_1-\Id\hat\mu_1}{\hat\sigma_1},\dots,\frac{\mub_p-\Id\hat\mu_p}{\hat\sigma_p}]\in\R^{n\times p}$, where $\hat\mu_i$ and $\hat\sigma_i$ denote the sample mean and sample standard deviation of $\mub_i$ and $\Id$ denotes the vector of ones in $\R^n$. The mean data matrix is defined as the row vector $\muhb = [\hat\mu_1,\dots,\hat\mu_p] \in \R^{1\times p}$. To reduce the effects of noise in the data set, each column of $\Xbtilde$ was smoothed using a Savitzky-Golay filter \cite{savitzky1964smoothing}. 

\begin{figure}[htbp]
		\centering
		\begin{minipage}[b]{\linewidth}
			\centering
			\subfloat{\includegraphics[width=0.45\linewidth]{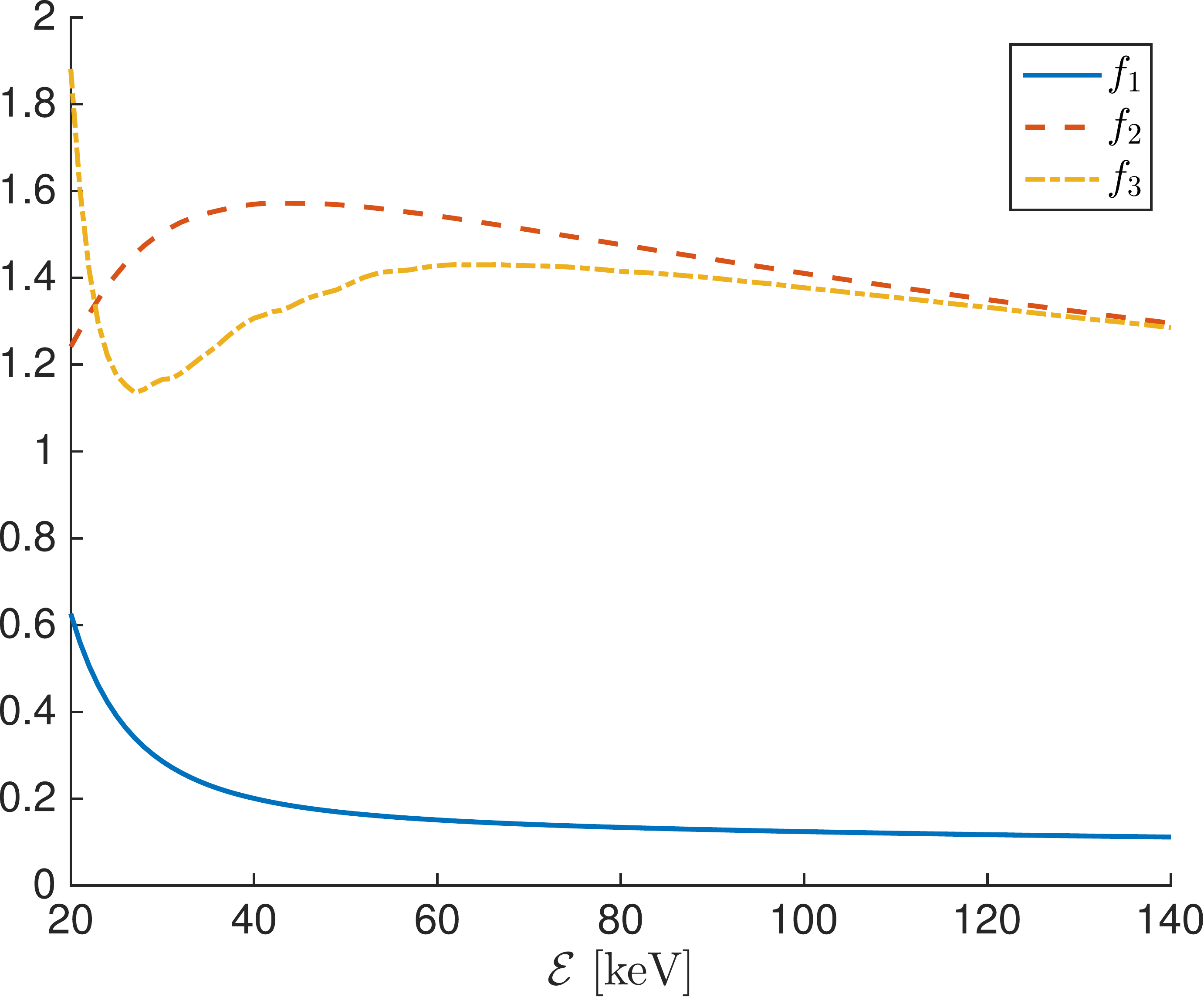}}							
		\end{minipage}
		\caption{Basis vectors obtained by the method described in Section \ref{sec:construction}.}\label{fig:basis_vectors}
	\end{figure}

	An $L$-dimensional, principal component representation of $\Xbtilde$ is found by computing its singular value decomposition (SVD),
	\begin{equation}
		\Xbtilde = \Ub\Sigb\Vb^T, \quad \Ub\in\R^{n\times n}, \; \Sigb \in \R^{n\times p}, \; \Vb\in\R^{p\times p},
	\end{equation}
	and letting the centered basis vectors $\fbtilde = [\fbtilde_1, \dots, \fbtilde_L]\in\R^{n\times L}$ be the first $L$ columns of $\Ub$. Since $\Ub$ is unitary it follows that $\fbtilde\,{}^T\!\fbtilde = \boldsymbol{I}_L$, i.e. the identity matrix of $\R^L$. Also, since $\Xbtilde{}^T\Id= \boldsymbol{0}_L$, i.e. the zero vector in $\R^L$, it holds that $\fbtilde\,{}^T\Id = \boldsymbol{0}_L$ if $L$ is smaller than or equal to the dimensionality of the data set $\Xb$, which we assume is the case. The basis coefficient vector $\ab_i \in \R^L$ of $\mub_i$ is given by the least-squares solution of
	\begin{equation}
		\fbtilde \ab_i  = \mub_i - \Id\hat\mu_i,
	\end{equation}
	which is given by
	\begin{equation}\label{eq:coefficient}
		\ab_i = \fbtilde\,{}^T(\mub_i - \Id\hat\mu_i) = \fbtilde\,{}^T\mub_i.
	\end{equation}
	Let $\ab = [\ab_1,\dots,\ab_p] = \fbtilde\,{}^T \Xb\in\R^{L\times p}$ be the matrix of basis coefficient vectors. The $L$-dimensional representation of $\Xb - \Id\muhb$ is given by $\fbtilde\ab = \fbtilde\fbtilde\,{}^T \Xb$. The $L$-dimensional representation of $\Xb$ is found by adding a mean component to each basis vector in $\fbtilde$. Guided by \eqref{eq:mean}, we take this mean component as the least-squares solution $\boldsymbol{\hat f} \in \R^{1\times L}$ of $\boldsymbol{\hat f}\ab = \muhb$, i.e.
	\begin{equation}
		\boldsymbol{\hat f} = \muhb \ab^T(\ab\ab^T)^{-1},
	\end{equation}
	and let $\boldsymbol{f} = \fbtilde + \Id \boldsymbol{\hat f}$. The obtained basis vectors are shown in Fig. \ref{fig:basis_vectors}. 

	\subsubsection{Observations on Optimality}

	Since the scale normalization of the data set $\Xb$ affects the obtained basis vectors, notions of optimality are somewhat arbitrary. However, to fix ideas, we show that for an already normalized $\Xb$, i.e. such that $\hat\sigma_i = 1$ for $i = 1,\dots,p$, the approximation error of any $L$-dimensional representation $\fb\ab$ of $\Xb$ is given by
	\begin{equation}
		\|\Xb - \fb\ab\|_2^2 = \|\Xbtilde - \fbtilde\ab\|_2^2 + n\|\muhb - \fbhat \ab \|_2^2. 
	\end{equation}
	\begin{proof}
	Let $\Xb\in\R^{n\times p}$ and let $\fb\ab$, where $\fb\in\R^{n\times L}$ and $\ab\in\R^{L\times p}$ be an $L$-dimensional representation of $\Xb$. Let further $\muhb\in\R^{1\times p}$ and $\fbhat\in\R^{1\times p}$ denote the row vectors of column-wise means of $\Xb$ and $\fb$. Lastly, let $\Xbtilde$ and $\fbtilde$ denote $\Xb - \Id\muhb$ and $\fb - \Id\fbhat$ respectively, where $\Id$ denotes the vector of ones in $\R^n$. The approximation error of $\fb\ab$ is then given by
	\begin{equation}
	\begin{split}
	&\|\Xb - \fb\ab\|_2^2 = \|\Xb - \Id\muhb + \Id\muhb - (\fb - \Id\fbhat + \Id\fbhat\,)\ab\|_2^2 \\
	=\;&\|\Xbtilde - \fbtilde\ab + \Id(\muhb - \fbhat \ab) \|_2^2 \\
	=\;&\|\Xbtilde - \fbtilde\ab\|_2^2 + \|\Id(\muhb - \fbhat \ab) \|_2^2 + 2(\Xbtilde - \fbtilde\ab)^T\Id(\muhb - \fbhat \ab) \\
	=\;&\|\Xbtilde - \fbtilde\ab\|_2^2 + n\|\muhb - \fbhat \ab \|_2^2,
	\end{split}
	\end{equation}
	where the last equality holds since $\Xbtilde{}^T\Id = \boldsymbol{0}$, $\fbtilde\,{}^T\Id = \boldsymbol{0}$ and $\Id^T\Id = n$.
	\end{proof}
	The centered basis vectors $\fbtilde$ (and the corresponding basis coefficients $\ab$) constructed from the SVD of $\Xbtilde$ can be found to be the minimizer of $\|\Xbtilde - \fbtilde\ab\|_2^2$ \cite{eckart1936approximation}, \cite{mirsky1960symmetric}. Given that choice of $\fbtilde$, the least-squares solution $\fbhat$ of $\|\muhb - \fbhat \ab \|_2^2$ clearly minimizes the total approximation error.

\subsection{Image Reconstruction}\label{sec:reconstruction}

	The optimization problem \eqref{eq:ML} was solved using a \textsc{Matlab} implementation of Newton's method as described in \cite[page 487]{boyd2009convex} for projection data generated from the simulated phantom described in Section \ref{sec:phantom}, using the basis vectors obtained by the method described in Section \ref{sec:construction}. Starting guesses of the projection line integrals $\Ab_i$ were obtained from a log-linearization of the parametrized forward model \eqref{eq:param_model} without a noise term, as in \cite{alvarez2013dimensionality}. These are given by the least-squares solution of the system
	\begin{equation}
		\Mb\Ab_i = \boldsymbol{c}_i,
	\end{equation}
	where the elements of $\Mb\in\R^{K\times L}$ and $\boldsymbol{c}_i \in \R^K$ are given by
	\begin{equation}
		M_{kl} = \frac{\int_\R w_k(\E)f_l(\E)d\E}{\int_\R w_k(\E)d\E}, \quad c_{ik} = -\ln \frac{y_{ik}}{\int_\R w_k(\E)d\E}.
	\end{equation}

\begin{figure}[htbp]
		\centering
		\begin{minipage}[b]{\linewidth}
			\centering
			\subfloat[First principal component basis image. \label{fig:basis_1}]{\includegraphics[width=0.45\linewidth]{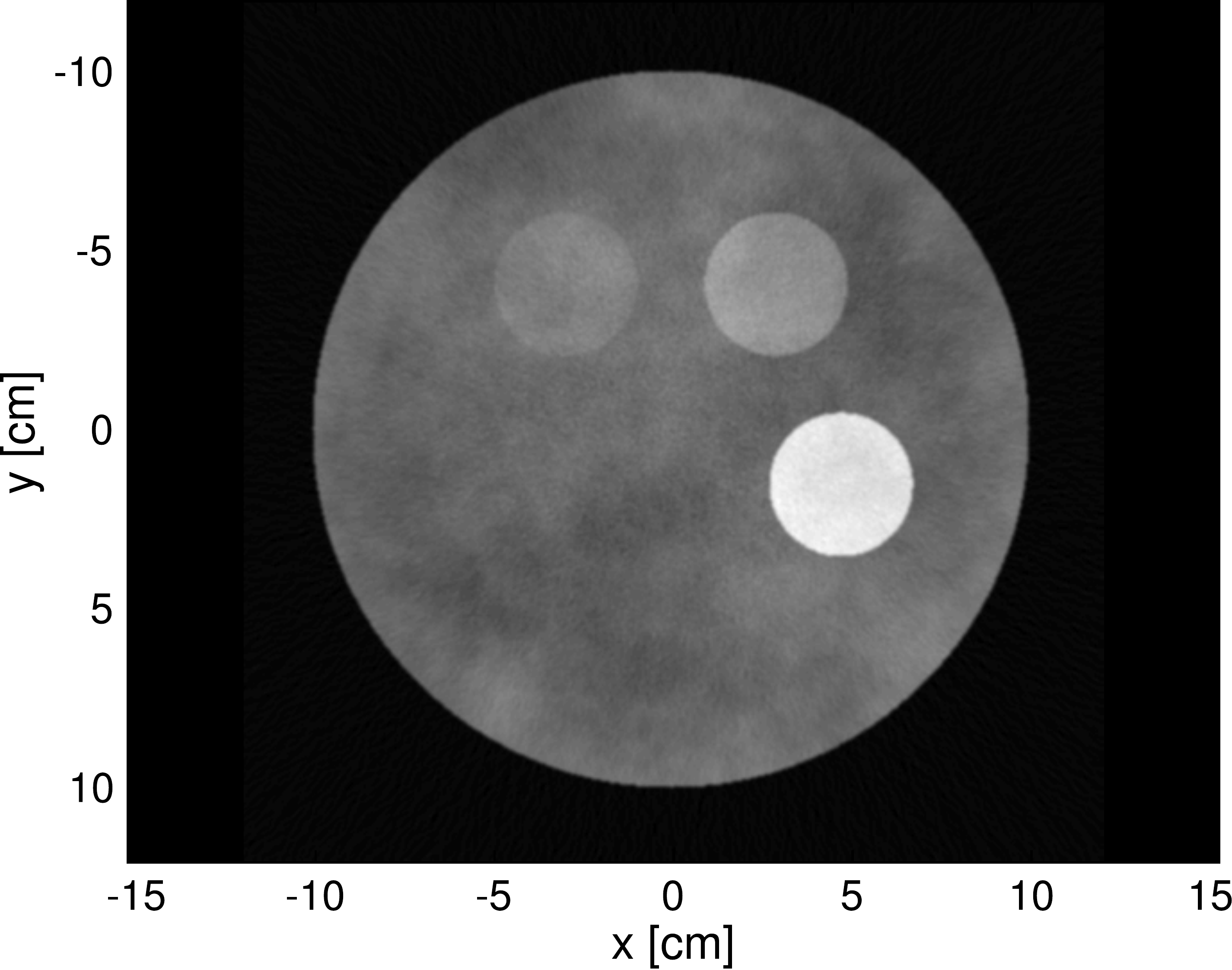}}				
			\hfil
			\subfloat[Second principal component basis image. \label{fig:basis_2}]{\includegraphics[width=0.45\linewidth]{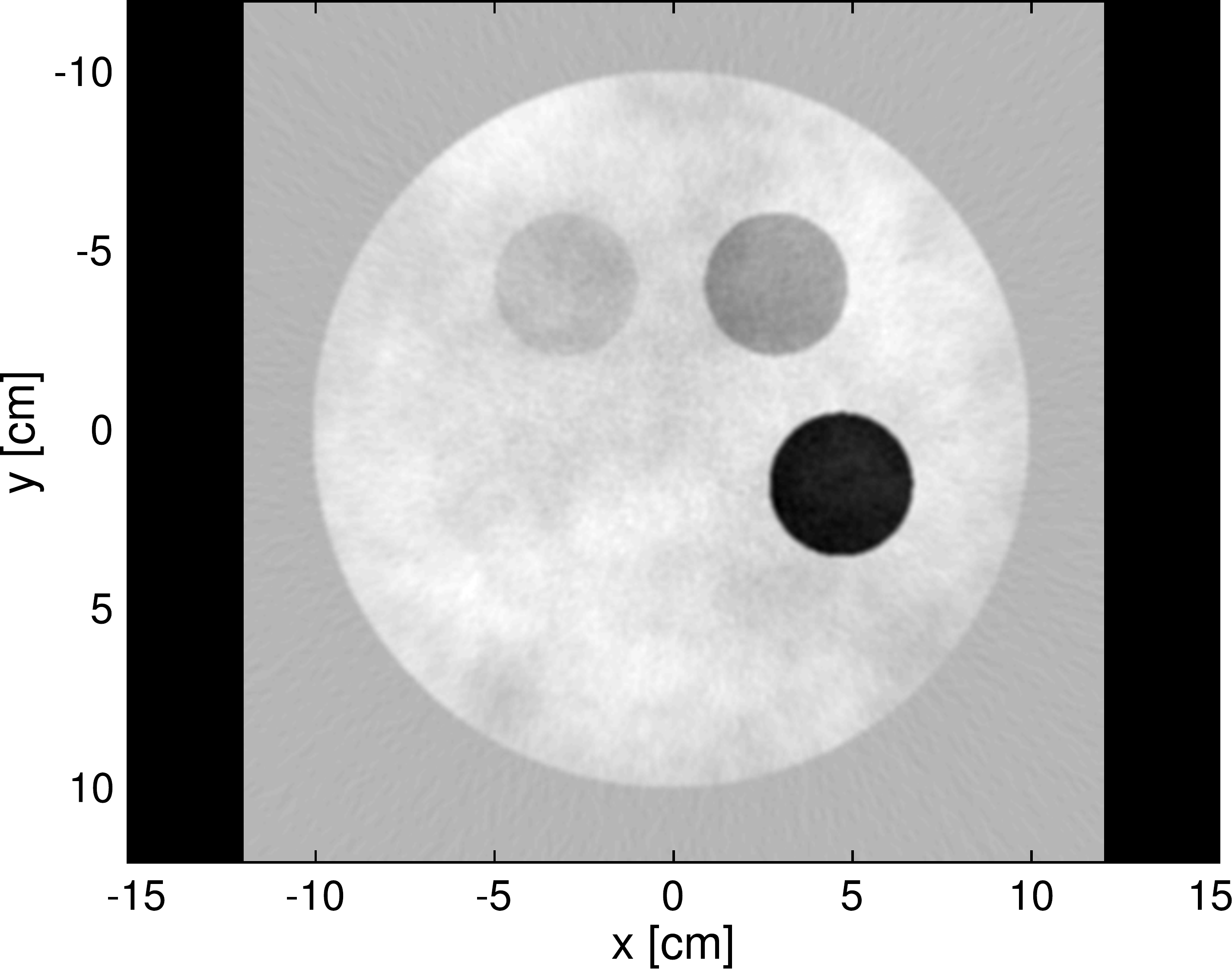}}	
			\hfil		
			\subfloat[Third principal component basis image. \label{fig:basis_3	}]{\includegraphics[width=0.45\linewidth]{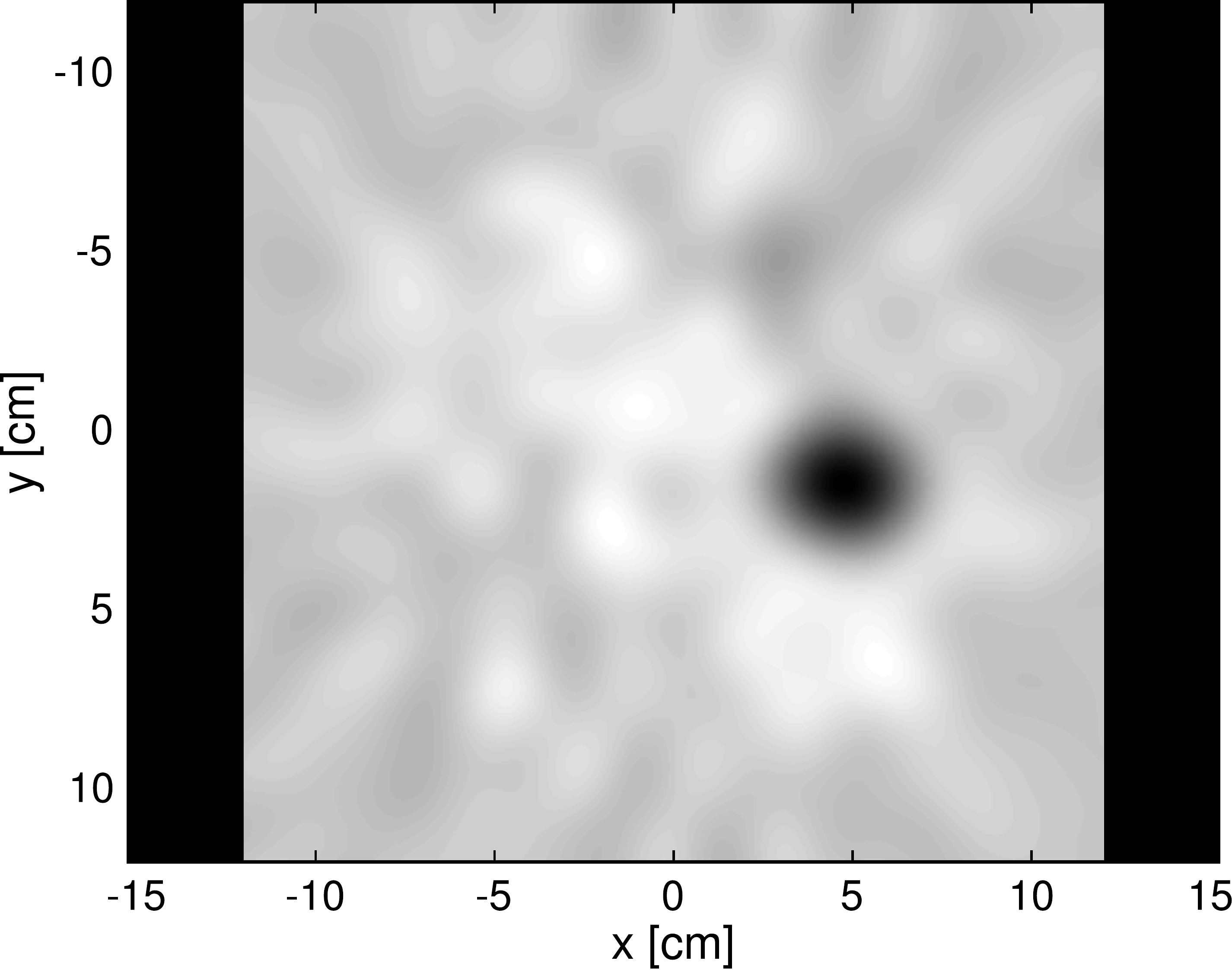}}			
		\end{minipage}
		\caption{Basis images obtained from $\tildeAbcal_3$ using FBP with filter frequencies given in Tab. \ref{tab:frequencies}.}\label{fig:basis_images}
	\end{figure}

	Let $\Abcal_2$ and $\Abcal_3$ denote the sets $\{\Ab_i^*\}_{i=1}^N$ obtained with two and three basis decompositions respectively. Furthermore, let $\tildeAbcal_3$ be the data set that consists of $\Abcal_2$ and the third component of $\Abcal_3$. The reason of considering also this data set is the expectation that the bias of the first two components of $\Abcal_2$ is low compared to the noise of the first two components of $\Abcal_3$. With photon-counting multibin systems it is possible to perform multiple basis decompositions and therefore also to combine data from the respective data sets.  Basis images were obtained from these data sets using \textsc{Matlab}'s implementation of filtered back-projection ({\tt iradon}) with the cosine filter. Filter frequencies were chosen for each basis image so as to minimize the mean squared error of each basis image over an ensemble of five phantoms. This may seem like an excessive optimization to the specific task, however, it can never completely remove the error due to the tissue variability that has been incorporated into the phantom and was therefore deemed to be justifiable. The chosen filter frequencies are presented in Tab. \ref{tab:frequencies}. The resulting images from $\tildeAbcal_3$ are shown in Fig. \ref{fig:basis_images}.
	
	\begin{table}[htbp]
		\centering
		\begin{tabular}{c | c | c | c}
		Data set & Frequency 1 & Frequency 2 & Frequency 3 \\
		\hline
		$\Abcal_2$ & 0.790 & 0.516 & \\
		$\tildeAbcal_3$ & 0.790 & 0.516 & 0.04 \\
		$\Abcal_3$ & 0.246 & 0.100 & 0.04 \\
		\end{tabular}\caption{FBP filter frequencies used to create basis images from the data sets $\Abcal_2$, $\Abcal_3$ and $\tildeAbcal_3$.}\label{tab:frequencies}
	\end{table}

	The background cancellation mappings of the type \eqref{eq:gamma2d} and \eqref{eq:gamma3d} were created using $\ab_{\text{liver}}$ and $\ab_{\text{adipose}}$ as the assumed coefficient vectors of the background tissue and $\ab_{\text{iron}}$ as the third material coefficient vector. These coefficient vectors were created using the relation \eqref{eq:coefficient}, with the centered basis vectors $\boldsymbol{\tilde{f}}$ obtained according to the description in Section \ref{sec:construction} and LAC data created using the recommended density and composition of elements to represent liver and adipose tissue given in the ICRU-44 \cite{white1989tissue}, with elemental LAC data from the XCOM database \cite{berger1998xcom}. The mixing constant $\rho$ of \eqref{eq:gamma2d} was assumend to be equal to one. The two-dimensional background cancellation mapping was applied to the images created from $\Abcal_2$ and the three-dimensional background cancellation mapping was applied to the ones created from both $\Abcal_3$ and $\tildeAbcal_3$. For validation purposes, the background cancellation mappings were also applied to basis images obtained directly from the phantom. The resulting estimation errors showed no dependence on the true iron weight map, consistent the derived estimation errors \eqref{eq:estimation_error_1} and \eqref{eq:estimation_error_2}.

\subsection{Figures of Merit}\label{sec:merit}

	To evaluate the performance of the background cancellation mappings, the mean squared error (MSE) of the estimated iron weight map is computed for each of the data sets $\Abcal_2$, $\Abcal_3$ and  $\tildeAbcal_3$. This procedure is repeated over an ensemble of five phantoms.



\section{Results}\label{sec:results}

\begin{figure}[htbp]
		\centering
		\begin{minipage}[b]{\linewidth}
			\centering
			\subfloat[Background cancellation image obtained from $\Abcal_2$. \label{fig:iron_1}]{\includegraphics[width=0.45\linewidth]{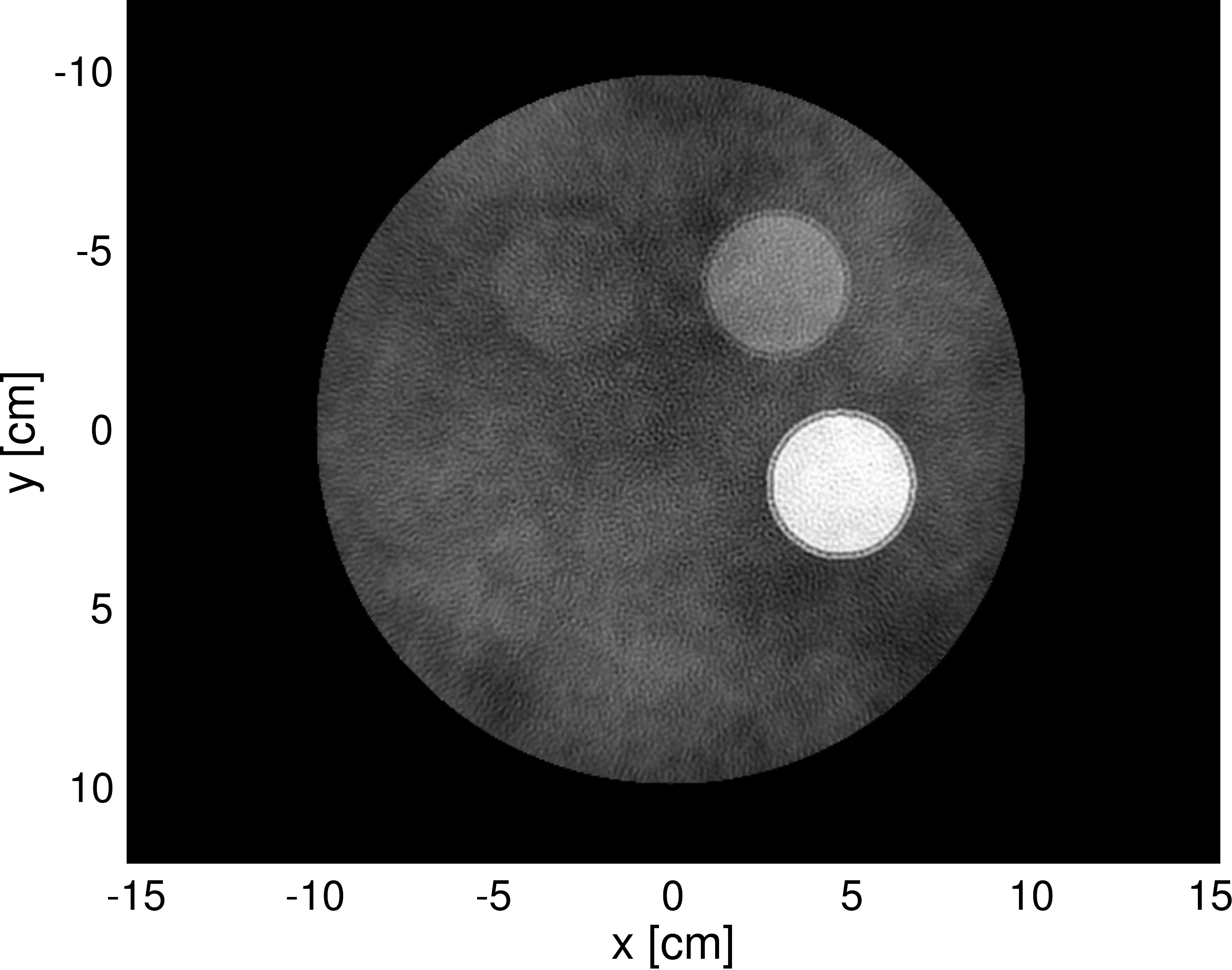}}				
			\hfil
			\subfloat[Background cancellation image obtained from $\Abcal_3$. \label{fig:iron_2}]{\includegraphics[width=0.45\linewidth]{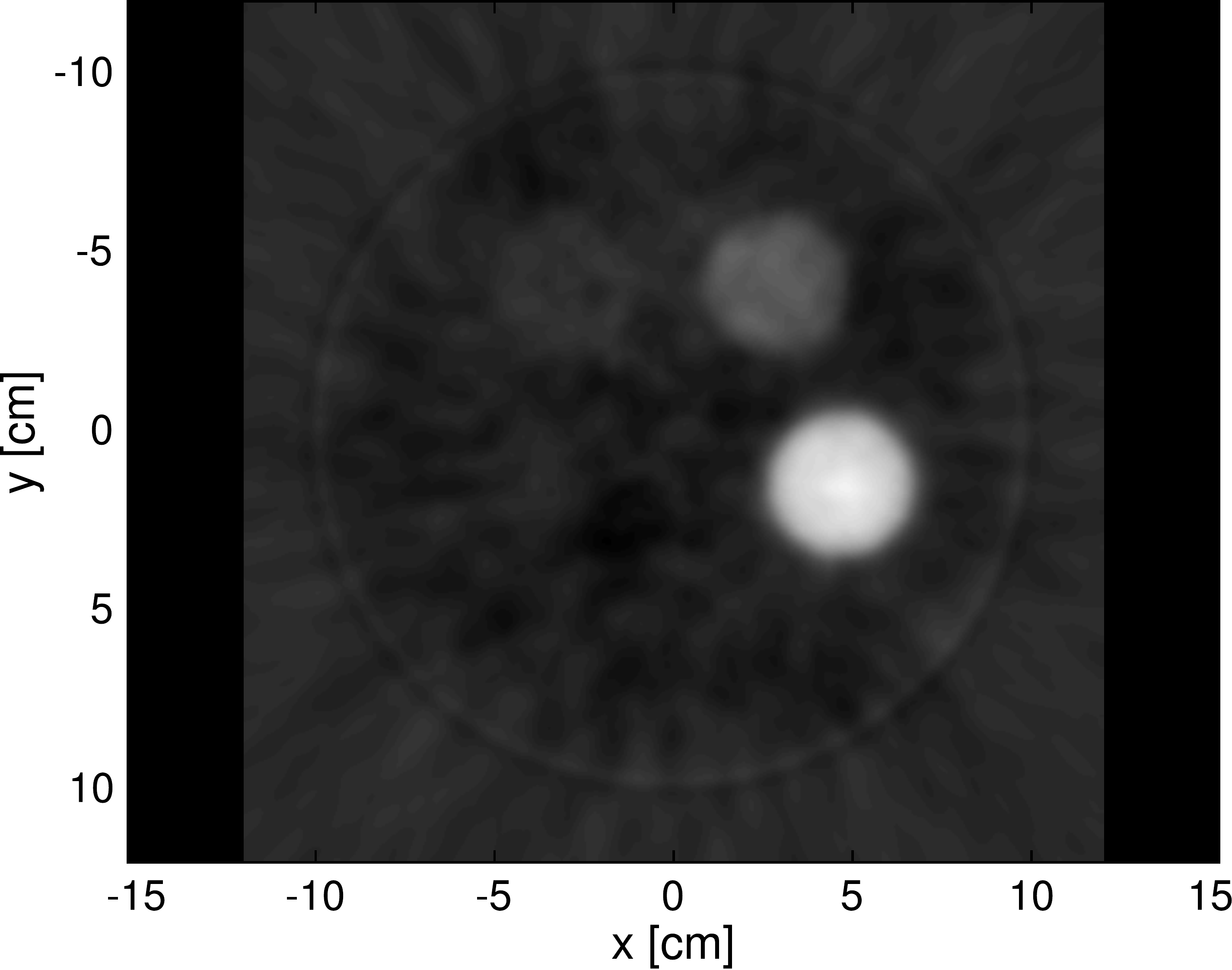}}	
			\hfil		
			\subfloat[Background cancellation image obtained from $\tildeAbcal_3$. \label{fig:iron_3}]{\includegraphics[width=0.45\linewidth]{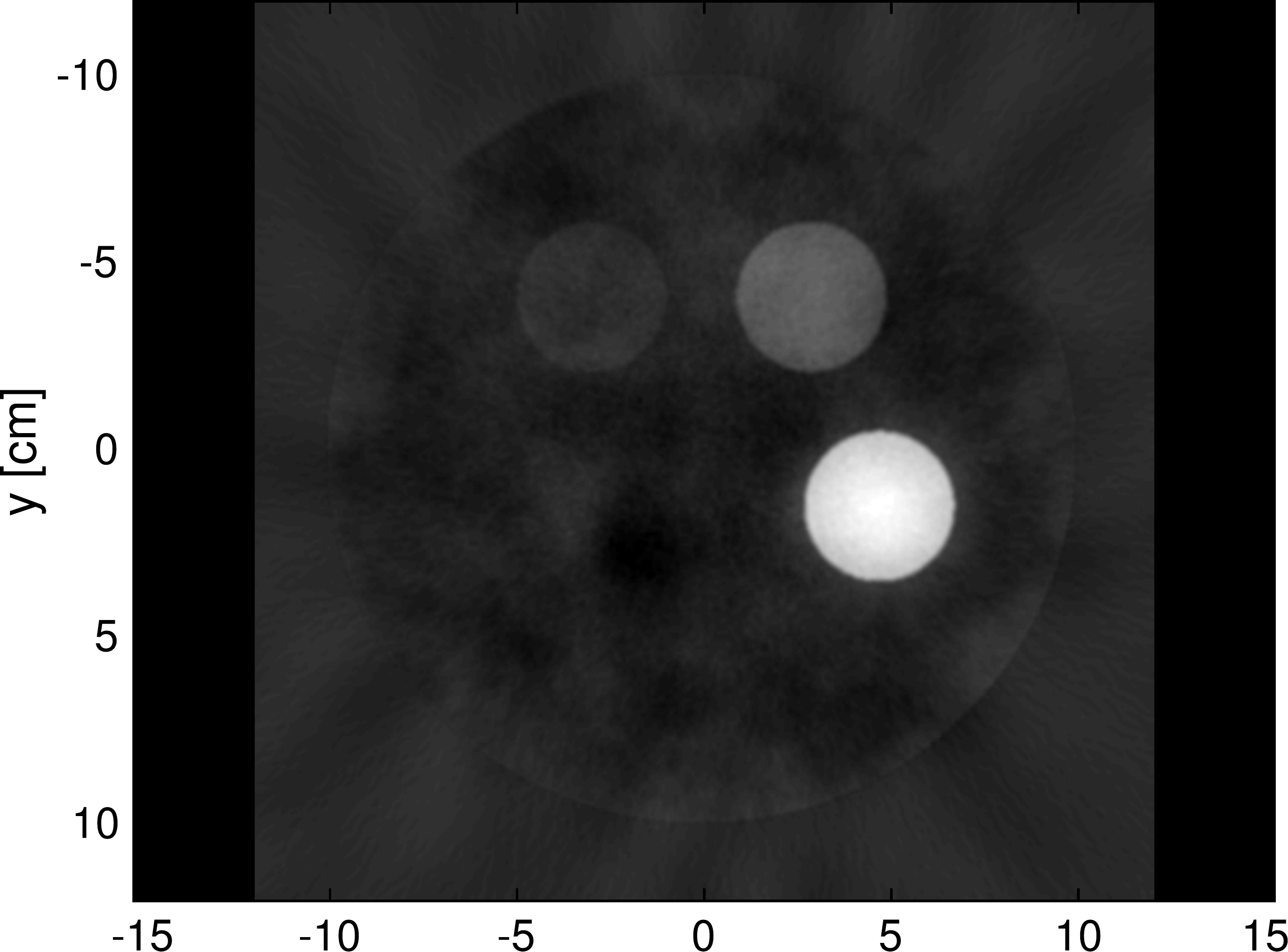}}			
		\end{minipage}
		\caption{Background cancellation images obtained from the data sets $\Abcal_2$, $\Abcal_3$ and $\tildeAbcal_3$.}\label{fig:iron_images}
	\end{figure}

	Background cancellation images created from the data sets $\Abcal_2$, $\Abcal_3$ and  $\tildeAbcal_3$ obtained from a single phantom are shown in Fig. \ref{fig:iron_images}. The resulting MSEs for each data set and each phantom in the simulated ensemble are presented in Tab. \ref{tab:MSEs}. 

	\begin{table}[htbp]
		\centering
		\begin{tabular}{c | c | c | c | c}
		Insert & $\text{MSE}(\Abcal_2)$ & $\text{MSE}(\Abcal_3)$ & $\text{MSE}(\tildeAbcal_3)$  & $\gamma$\\
		\hline
		1 & 1.20e-3 & 1.19e-7 & 7.74e-8 & 3.42e-3 \\
		2 & 1.19e-3 & 1.15e-7 & 7.86e-8 & 1.14e-3 \\
		3 & 1.22e-3 & 1.18e-7 & 7.87e-8 & 3.81e-4 \\
		4 & 1.16e-3 & 1.10e-7 & 7.18e-8 & 1.27e-4 \\
		5 & 1.09e-3 & 1.09e-7 & 7.36e-8 & 4.23e-5 \\
		\hline
		Mean MSE & 1.17e-3 & 1.14e-7 & 7.60e-8 & \\
		$\sqrt{\text{Mean MSE}}$ & 3.42e-2 & 3.38e-4 & 2.76e-4 &
		\end{tabular}
		\caption{Mean squared errors of the estimated iron weight maps $\hat\gamma$ for a set of five phantom realizations. For comparison, the rightmost column shows the true iron weight map values for each insert, in decreasing order.}\label{tab:MSEs}
	\end{table}

\section{Conclusion}\label{sec:conclusion}

 	The results show that in terms of quantitative information, i.e. a small MSE compared to the signal, the inclusion of a third basis function in the basis decomposition is essential for the studied imaging task, both in the case of a three-basis decomposition and the combination of two- and three-basis decompositions. It has not yet been clear that the use of a third basis function could be beneficial (in the absence of contrast agents), even though it has been established that the dimensionality of body material LACs is greater than two. In extension, the results suggest that the use of three basis functions is necessary for soft tissue background cancellation methods that are robust to inter-patient tissue variability.




\section*{Disclosure}
Fredrik Grönberg discloses past financial interests in Prismatic Sensors AB and is currently employed by GE Healthcare. Hans Bornefalk discloses past financial interests in Prismatic Sensors AB.  Mats U. Persson discloses past financial interests in Prismatic Sensors AB and research collaboration with GE Healthcare. 

\section*{References}
\bibliographystyle{spiejour}

\begin{thebibliography}{10}

\bibitem{alvarez1976energy}
R.~E. Alvarez and A.~Macovski, ``Energy-selective reconstructions in x-ray
  computerised tomography,'' {\em Physics in medicine and biology} {\bf 21}(5),
  733  (1976).

\bibitem{macovski1976energy}
A.~Macovski, R.~Alvarez, J.-H. Chan, {\em et~al.}, ``Energy dependent
  reconstruction in x-ray computerized tomography,'' {\em Computers in biology
  and medicine} {\bf 6}(4), 325--336  (1976).

\bibitem{lehmann1986energy}
L.~Lehmann and R.~Alvarez, ``Energy-selective radiography a review,'' in {\em
  Digital Radiography},  145--188, Springer  (1986).

\bibitem{gingold1992systematic}
E.~L. Gingold and B.~H. Hasegawa, ``Systematic bias in basis material
  decomposition applied to quantitative dual-energy x-ray imaging,'' {\em
  Medical physics} {\bf 19}(1), 25--33  (1992).

\bibitem{williamson2006two}
J.~F. Williamson, S.~Li, S.~Devic, {\em et~al.}, ``On two-parameter models of
  photon cross sections: application to dual-energy ct imaging,'' {\em Medical
  physics} {\bf 33}(11), 4115--4129  (2006).

\bibitem{bornefalk2012xcom}
H.~Bornefalk, ``Xcom intrinsic dimensionality for low-z elements at diagnostic
  energies,'' {\em Medical physics} {\bf 39}(2), 654--657  (2012).

\bibitem{berger1998xcom}
M.~J. Berger, J.~Hubbell, S.~Seltzer, {\em et~al.}, ``Xcom: photon cross
  sections database,'' {\em NIST Standard reference database} {\bf 8}(1),
  3587--3597  (1998).

\bibitem{alvarez2013dimensionality}
R.~E. Alvarez, ``Dimensionality and noise in energy selective x-ray imaging,''
  {\em Medical physics} {\bf 40}(11), 111909  (2013).

\bibitem{nielsen1992noninvasive}
P.~Nielsen, R.~Engelhardt, R.~Fischer, {\em et~al.}, ``Noninvasive liver-iron
  quantification by computed tomography in iron-overloaded rats,'' {\em
  Investigative radiology} {\bf 27}(4), 312--316  (1992).

\bibitem{jolliffe2005principal}
I.~Jolliffe, {\em Principal component analysis}, Wiley Online Library  (2005).

\bibitem{weaver1985attenuation}
J.~B. Weaver and A.~L. Huddleston, ``Attenuation coefficients of body tissues
  using principal-components analysis,'' {\em Medical physics} {\bf 12}(1),
  40--45  (1985).

\bibitem{gronberg2015third}
F.~Gr\"onberg, M.~Persson, and H.~Bornefalk, ``Third material separation in
  spectral ct with basis decomposition,'' in {\em Proc. Intl. Mtg. on Fully 3D
  Image Recon. in Rad. and Nuc. Med},   (2015).

\bibitem{roessl2009cramer}
E.~Roessl and C.~Herrmann, ``Cram{\'e}r--rao lower bound of basis image noise
  in multiple-energy x-ray imaging,'' {\em Physics in medicine and biology}
  {\bf 54}(5), 1307  (2009).

\bibitem{cranley1997catalogue}
K.~Cranley, B.~Gilmore, G.~Fogarty, {\em et~al.}, ``Catalogue of diagnostic
  x-ray spectra and other data,'' {\em IPEM report} {\bf 78}  (1997).

\bibitem{long2012multi}
Y.~Long and J.~A. Fessler, ``Multi-material decomposition using statistical
  image reconstruction in x-ray ct,'' {\em Proc. 2nd Int. Mtg. on image
  formation in X-ray CT} , 413--6  (2012).

\bibitem{schirra2013statistical}
C.~O. Schirra, E.~Roessl, T.~Koehler, {\em et~al.}, ``Statistical
  reconstruction of material decomposed data in spectral ct,'' {\em Medical
  Imaging, IEEE Transactions on} {\bf 32}(7), 1249--1257  (2013).

\bibitem{white1989tissue}
D.~White, J.~Booz, R.~Griffith, {\em et~al.}, ``Tissue substitutes in radiation
  dosimetry and measurement,'' {\em ICRU Report} {\bf 44}  (1989).

\bibitem{richmond1985icrp}
C.~Richmond, ``Icrp publication 23,'' {\em International Journal of Radiation
  Biology and Related Studies in Physics, Chemistry and Medicine} {\bf 48}(2),
  285--285  (1985).

\bibitem{fischer2004average}
H.~Fischer, I.~Polikarpov, and A.~F. Craievich, ``Average protein density is a
  molecular-weight-dependent function,'' {\em Protein Science} {\bf 13}(10),
  2825--2828  (2004).

\bibitem{durnin1974body}
J.~Durnin and J.~Womersley, ``Body fat assessed from total body density and its
  estimation from skinfold thickness: measurements on 481 men and women aged
  from 16 to 72 years,'' {\em British journal of nutrition} {\bf 32}(01),
  77--97  (1974).

\bibitem{miller1986definition}
G.~S. Miller, ``The definition and rendering of terrain maps,'' in {\em ACM
  SIGGRAPH Computer Graphics},   {\bf 20}(4), 39--48, ACM  (1986).

\bibitem{miranda2008new}
A.~A. Miranda, Y.-A. Le~Borgne, and G.~Bontempi, ``New routes from minimal
  approximation error to principal components,'' {\em Neural Processing
  Letters} {\bf 27}(3), 197--207  (2008).

\bibitem{savitzky1964smoothing}
A.~Savitzky and M.~J. Golay, ``Smoothing and differentiation of data by
  simplified least squares procedures.,'' {\em Analytical chemistry} {\bf
  36}(8), 1627--1639  (1964).

\bibitem{eckart1936approximation}
C.~Eckart and G.~Young, ``The approximation of one matrix by another of lower
  rank,'' {\em Psychometrika} {\bf 1}(3), 211--218  (1936).

\bibitem{mirsky1960symmetric}
L.~Mirsky, ``Symmetric gauge functions and unitarily invariant norms,''
  (1960).

\bibitem{boyd2009convex}
S.~Boyd and L.~Vandenberghe, {\em Convex optimization}, Cambridge university
  press  (2009).

\end{thebibliography}

\end{document}